\documentclass[final,5p,times]{elsarticle}
\usepackage{graphicx}
\usepackage{bm}
\usepackage{amssymb}

\journal{Physics Letters B}

\begin{document}

\def \apj  {ApJ}
\def \apjl  {ApJL}
\def \aj  {AJ}
\def\mnras {MNRAS}
\def \etal {{\it et~al.}}
\def \chisq  {\ifmmode  \chi^2   \else  $\chi^2$  \fi}  
\def \chisqr {\ifmmode \chi^2_{\rm r} \else $\chi^2_{\rm r}$ \fi}
\def \spose#1{\hbox  to 0pt{#1\hss}}  
\def \lta{\mathrel{\spose{\lower 3pt\hbox{$\sim$}}\raise  2.0pt\hbox{$<$}}}
\def \gta{\mathrel{\spose{\lower  3pt\hbox{$\sim$}}\raise 2.0pt\hbox{$>$}}}
\def \ion#1#2{#1{\footnotesize{#2}}\relax} 
\def \ha  {\ifmmode H\alpha \else H$\alpha $ \fi} 
\def \hi {\ion{H}{I}} 
\def \hii {\ion{H}{II}} 
\def \oii {[\ion{O}{II}]} 

\def \kmsmpc {\>{\rm km}\,{\rm s}^{-1}\,{\rm Mpc}^{-1}}
\def \kms {\ifmmode  \,\rm km\,s^{-1} \else $\,\rm km\,s^{-1}  $ \fi }
\def \kpc {\ifmmode  {\rm kpc}  \else ${\rm  kpc}$ \fi  }  
\def \Msun {\ifmmode M_{\odot} \else $M_{\odot}$ \fi} 
\def \Mearth {\ifmmode M_{\oplus} \else $M_{\oplus}$ \fi} 
\def \hMsun {\ifmmode h^{-1}\,\rm M_{\odot} \else $h^{-1}\,\rm M_{\odot}$ \fi}
\def \hhMsun {\ifmmode h^{-2}\,\rm M_{\odot}\else $h^{-2}\,\rm M_{\odot}$ \fi}
\def \Lsun {\ifmmode L_{\odot} \else $L_{\odot}$ \fi} 
\def \hhLsun {\ifmmode h^{-2}\,\rm L_{\odot} \else $h^{-2}\,\rm L_{\odot}$ \fi}

\def \LCDM {\ifmmode \Lambda{\rm CDM} \else $\Lambda{\rm CDM}$ \fi}
\def \sig8 {\ifmmode \sigma_8 \else $\sigma_8$ \fi} 
\def \OmegaM {\ifmmode \Omega_{\rm M} \else $\Omega_{\rm M}$ \fi} 
\def \OmegaL {\ifmmode \Omega_{\rm \Lambda} \else $\Omega_{\rm \Lambda}$\fi} 
\def \Deltavir {\ifmmode \Delta_{\rm vir} \else $\Delta_{\rm vir}$ \fi}
\def\lya{Lyman-$\alpha$}
\def\hang{\par\hangindent=\parindent\noindent}

\def \LCDM {\ifmmode \Lambda{\rm CDM} \else $\Lambda{\rm CDM}$\fi}
\def \lCDM {\ifmmode \Lambda{\rm CDM} \else $\Lambda{\rm CDM}$ \fi}
\def \lcdm {\ifmmode \Lambda{\rm CDM} \else $\Lambda{\rm CDM}$\fi}

\def \rs {\ifmmode r_{\rm s} \else $r_{\rm s}$ \fi} 
\def \rrm2 {\ifmmode r_{-2} \else $r_{-2}$ \fi} 
\def \ccm2 {\ifmmode c_{-2} \else$c_{-2}$ \fi} 
\def \cvir {\ifmmode c_{\rm vir} \else $c_{\rm vir}$ \fi} 
\def \cbar {\ifmmode \overline{c} \else $\overline{c}$ \fi}

\def \R200 {\ifmmode R_{200} \else $R_{200}$ \fi} 
\def \Rvir {\ifmmode R_{\rm vir} \else $R_{\rm vir}$ \fi}

\def \v200 {\ifmmode V_{200} \else $V_{200}$ \fi} 
\def \Vvir {\ifmmode V_{\rm  vir} \else  $V_{\rm vir}$  \fi} 
\def  \Vhalo  {\ifmmode V_{\rm halo} \else $V_{\rm halo}$ \fi}

\def \M200 {\ifmmode M_{200} \else $M_{200}$ \fi} 
\def \Mvir {\ifmmode M_{\rm  vir} \else $M_{\rm  vir}$ \fi}  
\def \Mshell  {\ifmmode M_{\rm shell} \else $M_{\rm shell}$ \fi}

\def \Nvir {\ifmmode N_{\rm  vir} \else $N_{\rm  vir}$ \fi}  

\def \Jvir {\ifmmode J_{\rm vir} \else $J_{\rm vir}$ \fi} 
\def \Jshell {\ifmmode J_{\rm shell} \else $J_{\rm shell}$ \fi}

\def \Evir {\ifmmode E_{\rm vir} \else $E_{\rm vir}$ \fi} 

\def \lam {\ifmmode \lambda  \else $\lambda$ \fi} 
\def \lamp {\ifmmode \lambda^{\prime} \else $\lambda^{\prime}$  \fi} 
\def \lampc {\ifmmode \lambda^{\prime}_{\rm c} \else
  $\lambda^{\prime}_{\rm c}$  \fi} 
\def \lambar {\ifmmode \bar{\lambda}  \else  $\bar{\lambda}$  \fi}  
\def  \lampbar  {\ifmmode \bar{\lambda^{\prime}} \else
  $\bar{\lambda^{\prime}}$\fi} 
\def \siglam {\ifmmode \sigma_{\lambda} \else $\sigma_{\lambda}$ \fi} 
\def \siglamp {\ifmmode                \sigma_{\lambda^{\prime}} \else
$\sigma_{\lambda^{\prime}}$\fi} 
\def  \sigl {\sigma_{\ln\lambda}} 
\def \sigc {\sigma_{\ln\cvir}}

\def \Rd {\ifmmode R_{\rm d} \else $R_{\rm d}$ \fi} 
\def \Rs {\ifmmode R_{\rm s} \else $R_{\rm s}$ \fi}  
\def \Rd {\ifmmode R_{\rm d} \else $R_{\rm d}$ \fi}  
\def \Rcool  {\ifmmode R_{\rm  cool}  \else $R_{\rm cool}$ \fi} 
\def \RIII {\ifmmode  3.2\Rs \else $3.2\Rs$ \fi} 
\def \RII {\ifmmode 2.2\Rs \else $2.2\Rs$  \fi} 
\def \Reff {\ifmmode R_{\rm eff} \else $R_{\rm  eff}$ \fi} 
\def  \rb {\ifmmode r_{\rm b}  \else $r_{\rm b}$ \fi}

\def  \Sigmacrit   {\ifmmode  \Sigma_{\rm  crit}   
\else  $\Sigma_{\rm crit}$\fi} 
\def \Sig0 {\ifmmode \Sigma_{0} \else $\Sigma_{0}$ \fi}

\def \muI {\ifmmode \mu_{0,I} \else $\mu_{0,I}$ \fi}

\def \mgal {\ifmmode m_{\rm gal} \else $m_{\rm gal}$ \fi} 
\def \md {\ifmmode m_{\rm d} \else $m_{\rm d}$ \fi} 
\def \ms {\ifmmode m_{\rm   s}   \else   $m_{\rm   s}$   \fi}   
\def   \mdbar   {\ifmmode {\overline{m}}_{\rm d} \else
  ${\overline{m}}_{\rm d}$ \fi} 
\def \msbar {\ifmmode  \bar{m}_{\rm  s}  \else  $\bar{m}_{\rm s}$
  \fi}  
\def  \Md {\ifmmode M_{\rm d}  \else $M_{\rm d}$ \fi} 
\def  \Ms {\ifmmode M_{\rm s} \else $M_{\rm  s}$ \fi} 
\def \Mb {\ifmmode  M_{\rm b} \else $M_{\rm b}$ \fi} 
\def \Mstar {\ifmmode  M_{\rm star} \else $M_{\rm star}$ \fi}
\def \Mdisk {\ifmmode M_{\rm disk} \else $M_{\rm disk}$ \fi}

\def \Jd {\ifmmode J_{\rm d} \else $J_{\rm d}$ \fi} 
\def \Jb {\ifmmode J_{\rm b} \else $J_{\rm b}$ \fi}  
\def \fb {\ifmmode  f_{\rm b} \else $f_{\rm b}$ \fi}

\def  \jd  {\ifmmode j_{\rm  d}  \else  $j_{\rm  d}$ \fi}  
\def  \jdmd {\ifmmode \frac{j_{\rm  d}}{m_{\rm d}} \else
  $\frac{j_{\rm d}}{m_{\rm d}}$ \fi} 
\def \fj {\ifmmode f_{\rm j} \else $f_{\rm j}$ \fi} 
\def \ft {\ifmmode f_{\rm t}  \else $f_{\rm t}$ \fi} 
\def  \fM {\ifmmode f_{\rm M} \else $f_{\rm M}$ \fi}

\def  \Vd {\ifmmode  V_{\rm  d}  \else $V_{\rm  d}$  \fi} 
\def  \Vcool {\ifmmode V_{\rm cool} \else $V_{\rm cool}$ \fi} 
\def \Vcirc {\ifmmode V_{\rm circ}  \else $V_{\rm circ}$  \fi} 
\def \VIII  {\ifmmode V_{3.2} \else $V_{3.2}$ \fi} 
\def  \VII {\ifmmode V_{2.2} \else $V_{2.2}$ \fi}
\def \Vobs {\ifmmode V_{\rm obs}  \else $V_{\rm obs}$ \fi} 
\def \Vdisk {\ifmmode V_{\rm disk} \else  $V_{\rm disk}$ \fi} 
\def \Vmax {\ifmmode V_{\rm  max} \else  $V_{\rm max}$  \fi} 
\def  \Vmaxobs{\ifmmode V_{\rm max}^{\rm obs}\else  $V_{\rm max}^{\rm
    obs}$\fi}  
\def \Vtot {\ifmmode V_{\rm tot} \else $V_{\rm tot}$  \fi} 
\def \Vrot {\ifmmode V_{\rm rot} \else  $V_{\rm rot}$  \fi} 
\def  \Vflat {\ifmmode  V_{\rm  flat} \else $V_{\rm flat}$ \fi}

\def \Ups {\ifmmode \Upsilon  \else $\Upsilon$ \fi} 
\def \YB {\ifmmode \Upsilon_B \else $\Upsilon_B$ \fi} 
\def \YI {\ifmmode  \Upsilon_I  \else $\Upsilon_I$ \fi} 
\def \DeltaIMF {\ifmmode \Delta_{\rm IMF} \else $\Delta_{\rm IMF}$ \fi}

\def\rhodrat{\rho_\mathrm{d}/\rho_\mathrm{h}}
\def\rhod{\rho_\mathrm{d}}
\def\rhoh{\rho_\mathrm{h}}
\def\GeV{GeV$/$c$^2$}
\def\TeV{TeV$/$c$^2$}
\def\mwimp{M_{\chi}}

\begin{frontmatter}

\title{Dark Matter Disc Enhanced Neutrino Fluxes from the Sun and Earth}

\author[uzh]{Tobias Bruch\corref{cor1}\fnref{fn1}}
\ead{tbruch@physik.uzh.ch}

\author[cal]{Annika H.\,G. Peter}
\ead{apeter@astro.caltech.edu}

\author[uzh]{Justin Read\fnref{fn2}}
\ead{justin@physik.uzh.ch}

\author[uzh]{Laura Baudis\fnref{fn1}}
\ead{lbaudis@physik.uzh.ch}

\author[uzh]{George Lake\fnref{fn2}}
\ead{lake@physik.uzh.ch}

\cortext[cor1]{Corresponding author : Phone: +41446355792; Fax: +41446355704}
\fntext[fn1]{Physics Institute}
\fntext[fn2]{Institute for Theoretical Physics}
\address[uzh]{University of Z\"urich, Winterthurerstrasse 190, CH-8057,  Z\"urich, Switzerland}
\address[cal]{California Institute of Technology, MS 105-24, CA 91125, Pasadena, USA}

\begin{abstract}
As disc galaxies form in a hierarchical cosmology, massive merging satellites are preferentially dragged towards the disc plane. The material accreted from these satellites forms a dark matter disc that contributes 0.25 - 1.5 times the non-rotating halo density at the solar position. Here, we show the importance of the dark disc for indirect dark matter detection in neutrino telescopes. Previous predictions of the neutrino flux from WIMP annihilation in the Earth and the Sun have assumed that Galactic dark matter is spherically distributed with a Gaussian velocity distribution, the standard halo model. Although the dark disc has a local density comparable to the dark halo, its higher phase space density at low velocities greatly enhances capture rates in the Sun and Earth. For typical dark disc properties, the resulting muon flux from the Earth is increased by three orders of magnitude over the SHM, while for the Sun the increase is an order of magnitude. This significantly increases the sensitivity of neutrino telescopes to fix or constrain parameters in WIMP models. The flux from the Earth is extremely sensitive to the detailed properties of the dark disc, while the flux from the Sun is more robust. The enhancement of the muon flux from the dark disc puts the search for WIMP annihilation in the Earth on the same level as the Sun for WIMP masses $\lesssim 100$\,GeV. 
\end{abstract}

\begin{keyword}
dark matter \sep cosmology \sep Galaxy formation \sep high energy astrophysics

\PACS 95.36.+d \sep 98.80.-k \sep 98.62.-g \sep 95.85.Ry \sep 98.70.Sa
\end{keyword}
\end{frontmatter}

\section{Introduction}
A mysterious dark matter makes up most of the mass in the Universe, providing us with a unique window into physics beyond the Standard Model. Among the many plausible dark matter particle candidates, Weakly Interacting Massive Particles (WIMPs) \cite{LW77, Gunn78, Ellis84} that arise in theories with supersymmetry (SUSY) \cite{Jungman} or universal extra dimensions (UED) \cite{Cheng02,Hooper07}, stand out as well-motivated and detectable. WIMPs may be detected directly by scattering in laboratory detectors \cite{GW85}, or indirectly by their annihilation products. The annihilation rate scales as the square of the WIMP density, so the most luminous sources are expected to be near the Galactic centre or the centre of dark matter sub-halos, where the dark matter density peaks \cite{Gunn78,SS84,Lake90,Kuhlen2008,Springel2008,Strigari2008}. In addition, the Sun and Earth capture WIMPs and may be seen as sources of WIMP annihilation \cite{Freese1986,Gould1987,Griest1987,Krauss1985,Silk1985}. In all cases, the annihilation rate is sensitive to the dark matter's phase space structure.

WIMPs can be gravitationally trapped inside the Sun and Earth by elastic scattering, if the final WIMP states have velocities below the escape velocity. To date, annihilation rates in the Sun and Earth have been estimated using the Standard Halo Model (SHM), which is modelled as a smooth, spherically symmetric density component with a non-rotating Gaussian velocity distribution (sometimes with an anisotropic velocity dispersion tensor). However, recent dark matter-only simulations show deviations from this picture in the form of many small amplitude density fluctuations on $\sim 100$ pc scales, a few large amplitude density fluctuations due to the presence of substructure (which makes up $\sim 0.5\%$ of the mass at the solar circle), and relatively small deviations of the velocity distribution from Gaussian \cite{Vogelsberger2008a,Zemp2008}. Even if such structures were to survive in the presence of a baryonic disc and bulge, the indirect detection signal from the Earth and Sun is unlikely to deviate significantly from the SHM prediction. This is because the annihilation rate is sensitive to the phase space density averaged over long ($\gtrsim 100\hbox{ Myr}$) timescales. As a result, indirect detection by annihilation in the Earth and Sun is only sensitive to the local dark matter {\it macrostructure}.

There is at least one local macro-structural component beyond the SHM, which has been discovered in recent simulations of galaxy formation that include baryons. The baryonic disc of the Milky Way draws satellites closer to the disc plane by dynamical friction, where they are disrupted by tides \cite{darkdisk1}. This results in a thick disc of dark matter with a mid-plane density of 0.25-1.5 times the local dark halo density and kinematics similar to the thick disc of stars \cite{Read08, Read09}. The dark disc boosts the flux in direct detection experiments at low energies and increases the annual modulation signal with an energy-dependent phase shift that will betray the mass of the dark matter particle \cite{darkdiskdd}.

In this Letter, we focus on the importance of the dark disc for the detection of neutrinos from WIMP annihilation in the cores of the Sun and Earth. WIMPs can annihilate into a wide range of final products, of which muon neutrinos can escape and reach terrestrial detectors. On Earth, these muon neutrinos produce muons in charged current interactions with nuclei: $\nu_{\mu}+N\rightarrow \mu^- +X$. The ultra-relativistic muons can be detected by their Cerenkov radiation in large water or ice-based neutrino telescopes. So far, neutrino telescopes have found no evidence for high-energy neutrinos of astrophysical origin above the detected atmospheric neutrino background. The most stringent bounds on high-energy neutrinos from the Sun and Earth come from Super-Kamiokande \cite{SuperK2004}, AMANDA \cite{amanda7yr} and IceCube \cite{icecube22}. Super-Kamiokande, a 50\,kt water Cerenkov detector, and AMANDA, located in the ice sheet at the Amundsen-Scott South pole station, have been taking data throughout the past decade. Their muon flux limits from the Earth and the Sun are of order $\Phi_\mu \sim 10^3-10^4\hbox{ km}^{-2}\hbox{ yr}^{-1}$ for energies $E_\mu > 1\hbox{ GeV}$, where the tighter constraints apply to higher WIMP masses. These flux limits and those shown in the figures below are based on the assumption of a hard WIMP annihilation spectrum. IceCube, currently under construction at the site of the AMANDA experiment, has achieved a similar sensitivity with even a small fraction ($\sim 27\%$) of the construction completed, and is expected to have a 5-year sensitivity to flux from the Sun of $\Phi_\mu \sim70$ km$^{-2} \hbox{ yr}^{-1}$  above a WIMP mass of $\sim 200$\,GeV. The expected 5-year sensitivity to flux from the Earth is $\Phi_\mu \sim20$ km$^{-2} \hbox{ yr}^{-1}$.

In this Letter, we show that the presence of a dark disc dramatically increases the parameter space of WIMP models to which neutrino telescopes are sensitive. In \S\ref{sec:dd} we introduce the dark disc and its properties used in this Letter. In \S\ref{sec:cap}, we derive the capture and annihilation rates of WIMPs in the Sun and Earth. In \S\ref{sec:dm}, we describe a particular WIMP candidate -- the neutralino in the Constrained Minimal Supersymmetric Model (CMSSM). In \S\ref{sec:results}, we calculate the expected muon flux for CMSSM neutralinos that are consistent with both astrophysical and collider constraints, and we show how the flux changes once we include the dark disc. In \S\ref{sec:discussion}, we discuss how our results are affected by uncertainties in the dark disc phase space density. Finally, in \S\ref{sec:conclusions}, we present our conclusions.

\section{The dark disc}\label{sec:dd}
As disc galaxies form in a hierarchical cosmology, many smaller satellite galaxies are accreted and dissolve. \citet{Read08} looked at the interaction of a disc with a cosmological distribution of satellites. They found that low inclination massive satellites were preferentially dragged into the disc plane by dynamical friction. The material accreted from these satellites settled into a thick disc of dark matter. \citet{Read09} studied {\it ab initio} simulations of galaxy formation where a thick dark disc was also found. In both studies, the dark disc had a density $\rhod$ that is $\sim 0.25 - 1.5$ times the local density of dark matter in the SHM $\rhoh$, a rotation lag with respect to the stellar disc of $|\mathbf{v}_\odot| \sim 0 - 150$\,km/s, and a near-isotropic 1D dispersion of $\sigma \sim 50-90$\,km/s. The uncertainties in these numbers reflect the unknown stochastic merger history of our Galaxy. 

An accreted thick stellar disc forms concurrent with the dark disc and shares similar kinematics. Although the origin of the Milky Way thick disc of stars remains under investigation, it has kinematic properties remarkably similar to the median dark disc distribution with $|\mathbf{v}_\odot| = 40-50$\,km/s with respect to the local circular velocity, and dispersions of $(\sigma_R,\sigma_\phi,\sigma_z) = (63,39,39)$\,km/s \cite{Read08, Read09}. For our present study, we model the distribution functions of both the SHM and dark disc as Gaussian,
\begin{eqnarray}
	f(\mathbf{u}) = \frac{1}{(2\pi \sigma^2)^{3/2}} \frac{\rho}{M_\chi} e^{-(\mathbf{u} + \mathbf{v}_\odot)^2/2\sigma^2}, \label{eq:gauss}
\end{eqnarray}
where $\mathbf{u}$ is the heliocentric WIMP speed, $\sigma$ is the 1D velocity dispersion and $\mathbf{v}_\odot$ is the lag between the dark matter particles and the Sun. $\rho$ is the WIMP density at the solar circle and $M_\chi$ is the WIMP mass.  For the SHM, $|\mathbf{v}_\odot| = 220$\,km/s and $\sigma = |\mathbf{v}_\odot| / \sqrt{2}$.  For the dark disc, we assume fiducial values of: $\rhod = \rhoh$ and $|\mathbf{v}_\odot| = \sigma = 50$\,km/s, consistent both with the median of the ranges found in \cite{Read08} and \cite{Read09}, and with the kinematics of the Milky Way thick disc stars.  The sensitivity of our results to these parameters is discussed in \S\ref{sec:discussion}.

\section{Capture from the dark disc}\label{sec:cap}
The capture rate from a nuclear species $i$ per unit volume shell of a celestial body is given by \cite{Gould1987}
\begin{equation}
\frac{dC_i}{dV} = \int_0^{u_{max}} du \int d\Omega_w f(\mathbf{u}) u w^2 \sigma_i n_i,
\end{equation}
\noindent
where $f(\mathbf{u})$ is the velocity distribution normalised such that $\int f(\mathbf{u}) d^3\mathbf{u} = \rho/M_{\chi}$. The velocity $w$ at a given shell is related to the velocity at infinity $u$ and the escape velocity $v$ at the shell by $w=\sqrt{u^2+v^2}$. The WIMP-nucleus cross section is $\sigma_i$, and $n_i$ is the number density of nuclear species $i$. The upper limit of the integration is
\begin{equation}
 u_{max}=2\frac{\sqrt{M_\chi m_i}}{M_\chi - m_i}v , \label{eq:umax}
\end{equation}
\noindent
where $m_i$ is the mass of nuclear species $i$. This ensures that only WIMPs that can scatter to a velocity lower than the escape velocity $v$ are included. For fixed mass $m_i$ and escape velocity $v$ this upper cut off decreases with increasing $\mwimp$. 

The annihilation rate per unit volume of WIMPs in the body is given by
\begin{eqnarray}
	\frac{d\Gamma_A}{dV} = \langle \sigma_A v \rangle_0 n_\chi^2(t,\mathbf{x}),
\end{eqnarray}
where $\langle \sigma_A v \rangle_0$ is the velocity-averaged annihilation cross section in the limit of non-relativistic speeds, and $n_\chi(t,\mathbf{x})$ is the number density of WIMPs in the body.  If WIMPs quickly thermalize with nuclei in the body once captured, the number density of WIMPs in that body can be described by
\begin{eqnarray}
	n_\chi(t,\mathbf{x}) = N_\chi(t) \tilde{n}_\chi(\mathbf{x}),
\end{eqnarray}
where $\int dV \tilde{n}_\chi(\mathbf{x}) = 1$.  In that case, the number $N_\chi(t)$ of WIMPs in the body is given by the solution to
\begin{eqnarray}\label{eq:dotN}
	\dot{N_\chi}(t) = C - 2 ~\Gamma_A
\end{eqnarray}
if $M_\chi \gg m_i$, where the total capture rate is $C = \sum_i C_i$ \cite{Spergel1985, Griest1987}.  The factor of 2 in Eq. \ref{eq:dotN} reflects the fact that for self-annihilating particles, two WIMPs are destroyed in each annihilation.  If the total capture rate $C$ is constant with time, the annihilation rate is given by
\begin{eqnarray}
	\Gamma_A = \frac{C}{2}\tanh^2(t/\tau),
\end{eqnarray}
with the equilibrium time $\tau$ given by
\begin{eqnarray}
	\tau = (CC_A)^{-1/2},~C_A = 2\langle \sigma v\rangle_0 \int dV \tilde{n}_\chi^2(\mathbf{x}).
\end{eqnarray}
For a WIMP of $M_\chi \sim 100\hbox{ GeV}$ with purely spin-independent interactions with baryons, $C$ is about nine orders of magnitude greater for the Sun than for the Earth, while $C_A$ is about three orders of magnitude smaller \cite{Griest1987}. Thus, equilibrium timescales tend to be orders of magnitude shorter in the Sun than in the Earth.  If the spin-independent WIMP-proton cross section is $\sigma_p^{SI} = 10^{-43}\hbox{ cm}^2$, $\tau \sim 10^8 \hbox{ yr}$ for the Sun and $\tau \sim 10^{11}$ yr for the Earth \cite{Jungman}.  Therefore, if the age of the Solar system is $t_\odot \approx 4.5 \hbox{ Gyr}$, $t_\odot/\tau \gg 1$ in the Sun and $\Gamma_A=C/2$ and is constant. For this set of WIMP parameters, WIMP annihilation will have reached equilibrium in the Sun. However, in the Earth, $t_\odot/\tau \ll 1$, and $\Gamma_A \propto C^2$ and is growing with time. For some of the models that we consider, the interaction rate becomes low enough that equilibrium is broken in the Sun also. However, these models appear near the bottoms of the plots, far away from the interesting detection thresholds.  

\begin{figure*}[!htb]
\begin{center}
 \includegraphics[width=0.45\textwidth]{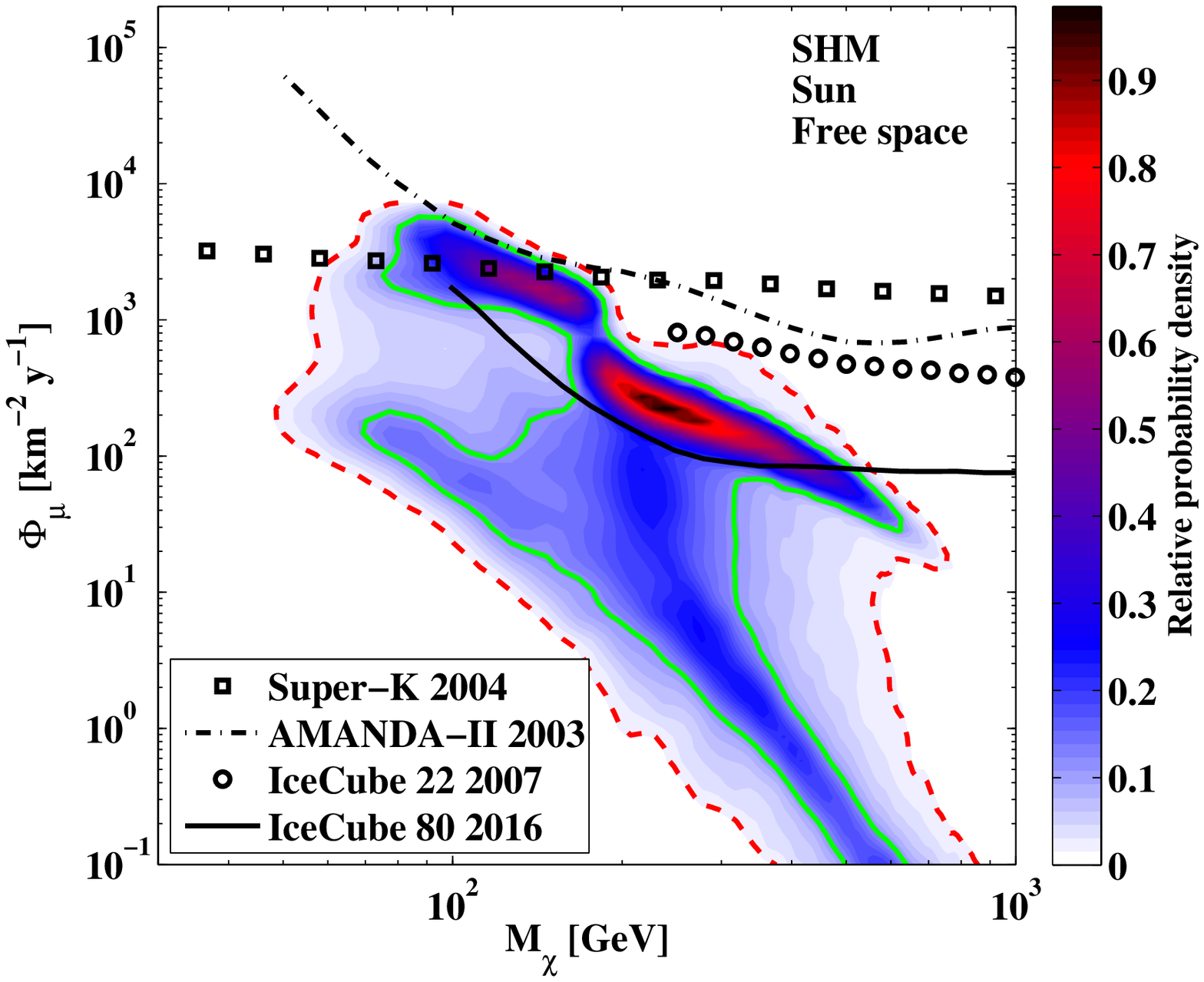} 
 \includegraphics[width=0.45\textwidth]{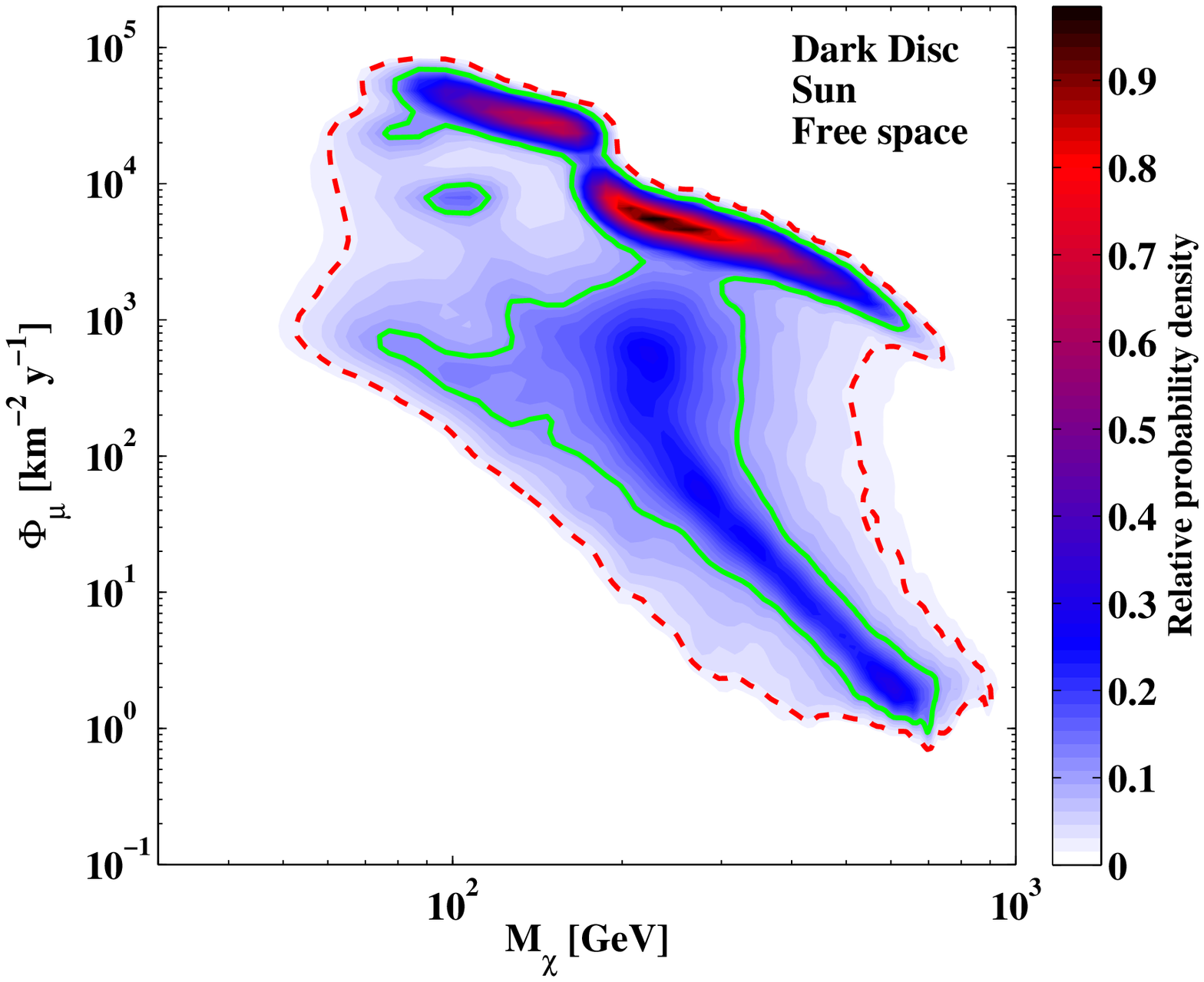} 
\label{fig:sunflux}
\caption{Muon flux $\Phi_{\mu}$ for $E_{\mu} > 1$\,GeV at the Earth's surface as a function of $\mwimp$ from neutrinos originating in the Sun, for the SHM (left panel) and the dark disc (right panel). The dark disc boosts the muon flux by an order of magnitude for $\rhodrat=1$. Current experimental constraints on the muon flux from the Sun from Super-Kamiokande \cite{SuperK2004}, AMANDA-II \cite{amanda7yr, amandab10} and IceCube22 \cite{icecube22} along with the expected sensitivity of IceCube80 are over-plotted on the left panel. The closed contours show -- 95\% (red/dashed) and 68\% (green/solid) -- of the probability density of CMSSM models consistent with both astrophysical and collider constraints, and assuming flat priors. The colour-bar gives the relative probability density (see \S\ref{sec:dm} for details).}
\end{center}
\end{figure*}

\begin{figure*}[!htb]
\begin{center}
 \includegraphics[width=0.45\textwidth]{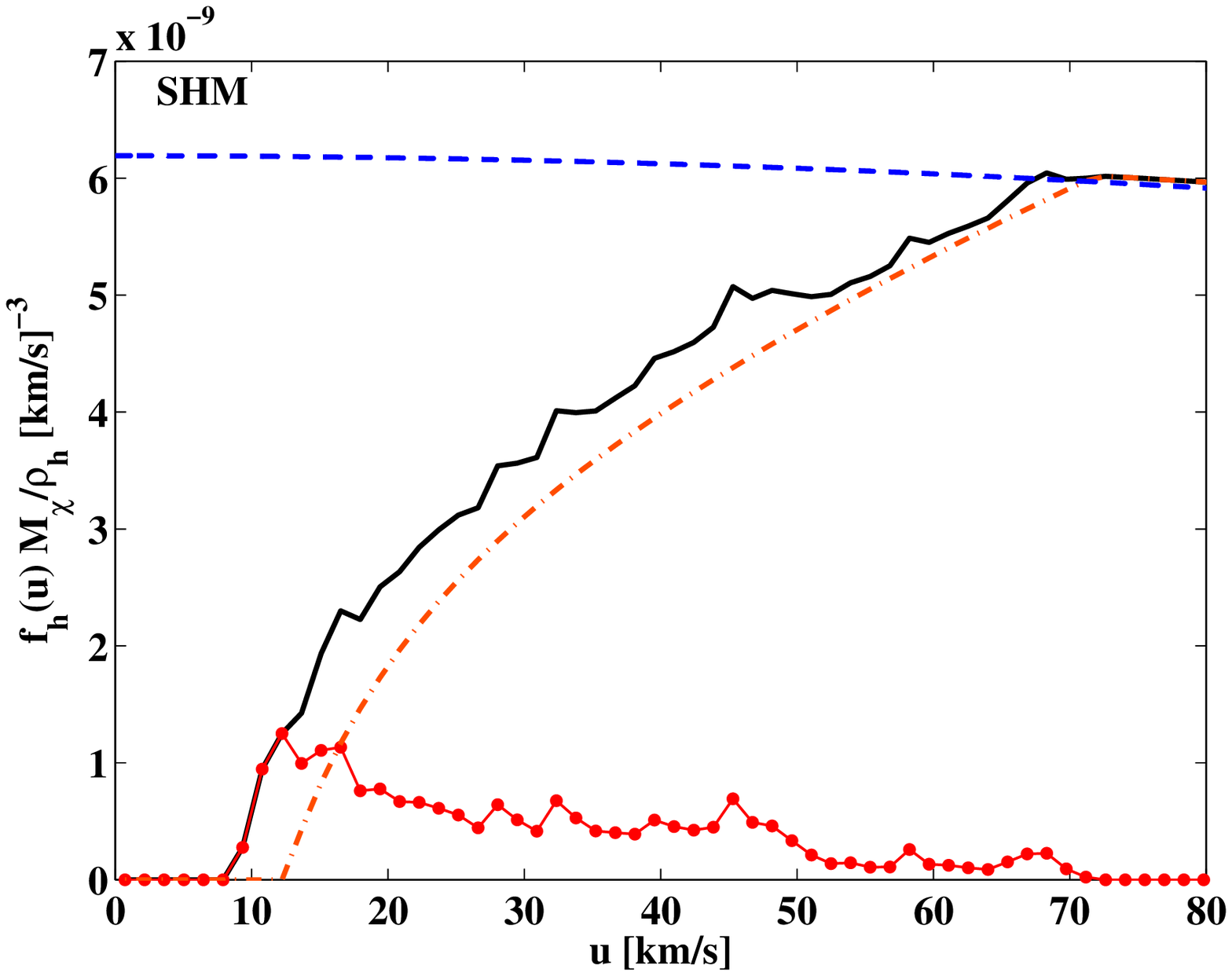} 
 \includegraphics[width=0.45\textwidth]{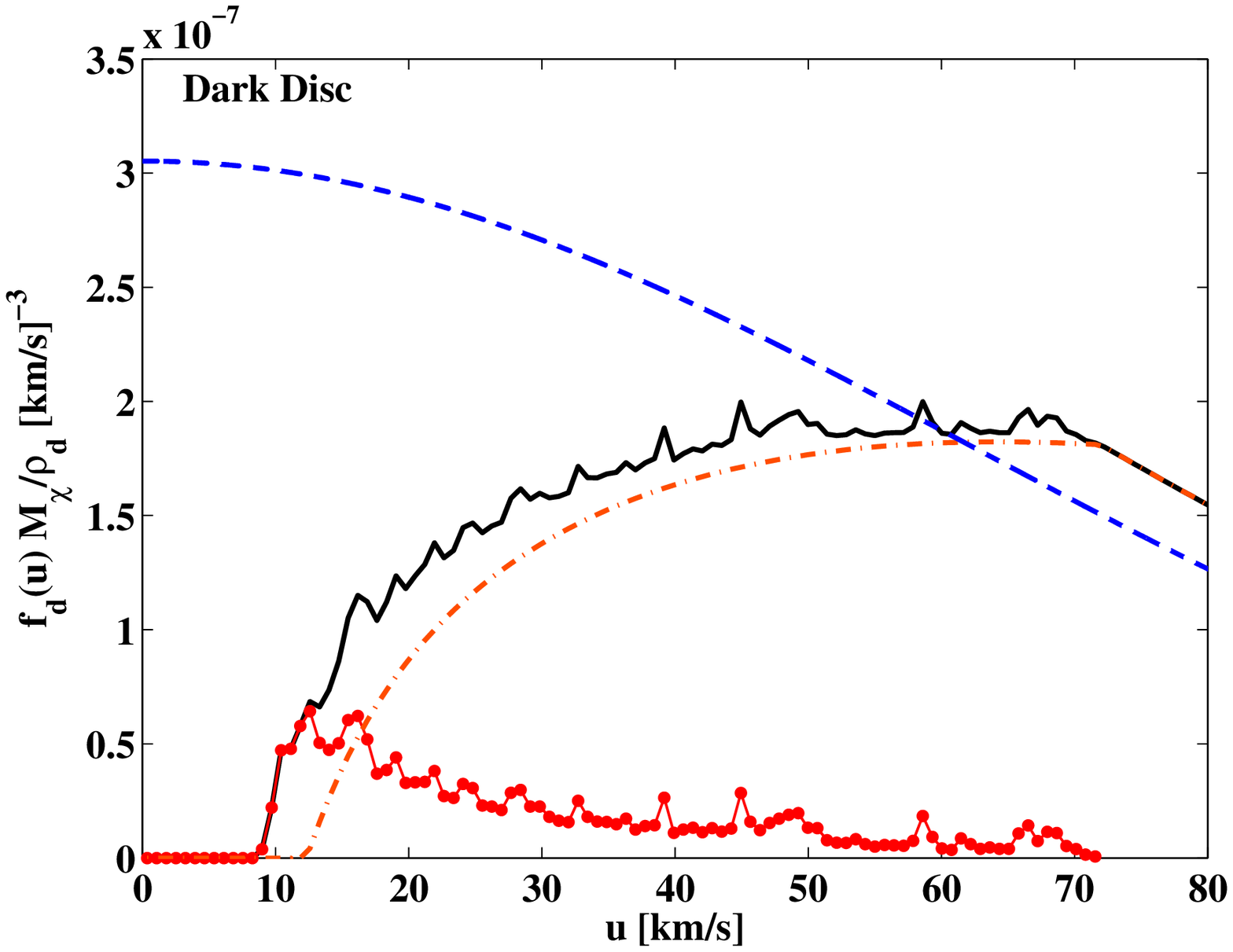} 
\label{fig:velodist}
\end{center}
\caption{The phase space density at low velocities for the SHM (left panel) and the dark disc (right panel). The black/solid curve is the summed distribution of bound (red/dotted) and unbound (orange/dash-dotted) particles from the Solar system simulations, used in the calculation of the capture rates. The blue/dashed line shows the distribution of the free space Gaussian approximation. \textit{Note the vertical scales of the two plots differ by two orders of magnitude.}}
\end{figure*}

\section{The dark matter candidate}\label{sec:dm}
For detailed calculations of capture and annihilation rates, we must assume a specific WIMP model for the particle's mass, scattering cross sections and annihilation channels. In supersymmetric extensions of the Standard Model with conserved R-parity, the lightest supersymmetric particle is a natural WIMP candidate. We choose the lightest neutralino in the CMSSM as the dark matter particle. The CMSSM reduces the free parameters in supersymmetry to three parameters at a gauge unification scale: the gaugino mass $m_{1/2}$, the scalar mass $m_{0}$ and the tri-linear coupling $A_0$. At the electroweak scale, the ratio of the Higgs vacuum expectation values tan$(\beta)$ and the sign of the Higgs/higgsino mass parameter $\mu$ are selected. Only the square of $\mu$ is calculated from the potential minimisation conditions of electroweak symmetry breaking. We scanned the CMSSM parameter space in the range of $m_{1/2}\in[0~4]$\,TeV, $m_0\in[0~4]$\,TeV, $A_0\in[-7~7]$\,TeV and tan$(\beta) \in [20~65]$ for $\mu>0$. The allowed parameter space, consistent with current experimental constraints is found with the publicly available SuperBayes package \cite{SuperBayes}. This Markov Chain Monte Carlo (MCMC) algorithm calculates the Bayesian posterior probability at each parameter point from the compatibility of the theoretical predictions with experimental constraints. The most relevant experimental constraints used in the MCMC are the cosmologically allowed relic density measured by the Wilkinson Microwave Anisotropy Probe (WMAP), electroweak precision observables and limits on the Higgs and lightest neutralino mass from colliders. All constraints listed in \citet{SuperBayes} have been used, except that the value for the dark matter density has been updated to the WMAP 5 year data release value, $\Omega_{DM} h^2 = 0.1099\pm 0.0062$, where $h$ is the Hubble constant in units of $100 h$\,km/s/Mpc and $h = 0.719$ is their best fit value \cite{WMAP2008}. Flat priors are used in the calculation of the Bayesian probability. Although the resulting allowed parameter space depends on our choice of priors, we are interested in the relative change in flux caused by the addition of the dark matter disc and thus our choice of priors is not important. The chains used in this study contain a total of $0.9 \times 10^6$ samples. The posterior probability density functions presented below are normalised to their maximum values, and \textit{not} to a total probability of 1. Accordingly we label these as ``relative probability densities''. An interface of the SuperBayes MCMC algorithm to DarkSusy (v5.03) \cite{DarkSusy} routines is used to calculate the capture rate and the resulting muon flux at the Earth.

\section{Neutrino flux from the Sun and Earth}\label{sec:results}
\subsection{The Sun}
In Fig. 1 we show the muon flux with an energy above 1\,GeV at the Earth resulting from WIMP capture and annihilation in the Sun. We show the flux as a function of $\mwimp$ for the SHM (left) and the dark disc (right) for $\rhodrat=1$ and $\rho_h = 0.3$\, GeV/cm$^{3}$.  The higher phase space density at low velocities for the dark disc strongly enhances the capture rate and hence the resulting muon flux at the detector site.  The flux expected from the dark disc is larger by approximately an order of magnitude (depending on the specific model) compared to the flux expected from the SHM, since the capture rate increases by approximately an order of magnitude, and $t_\odot / \tau \gg 1$ for most of the models in the figure (see \S\ref{sec:cap}). 

\begin{figure*}[!htb]
\begin{center}
 \includegraphics*[width=0.45\textwidth]{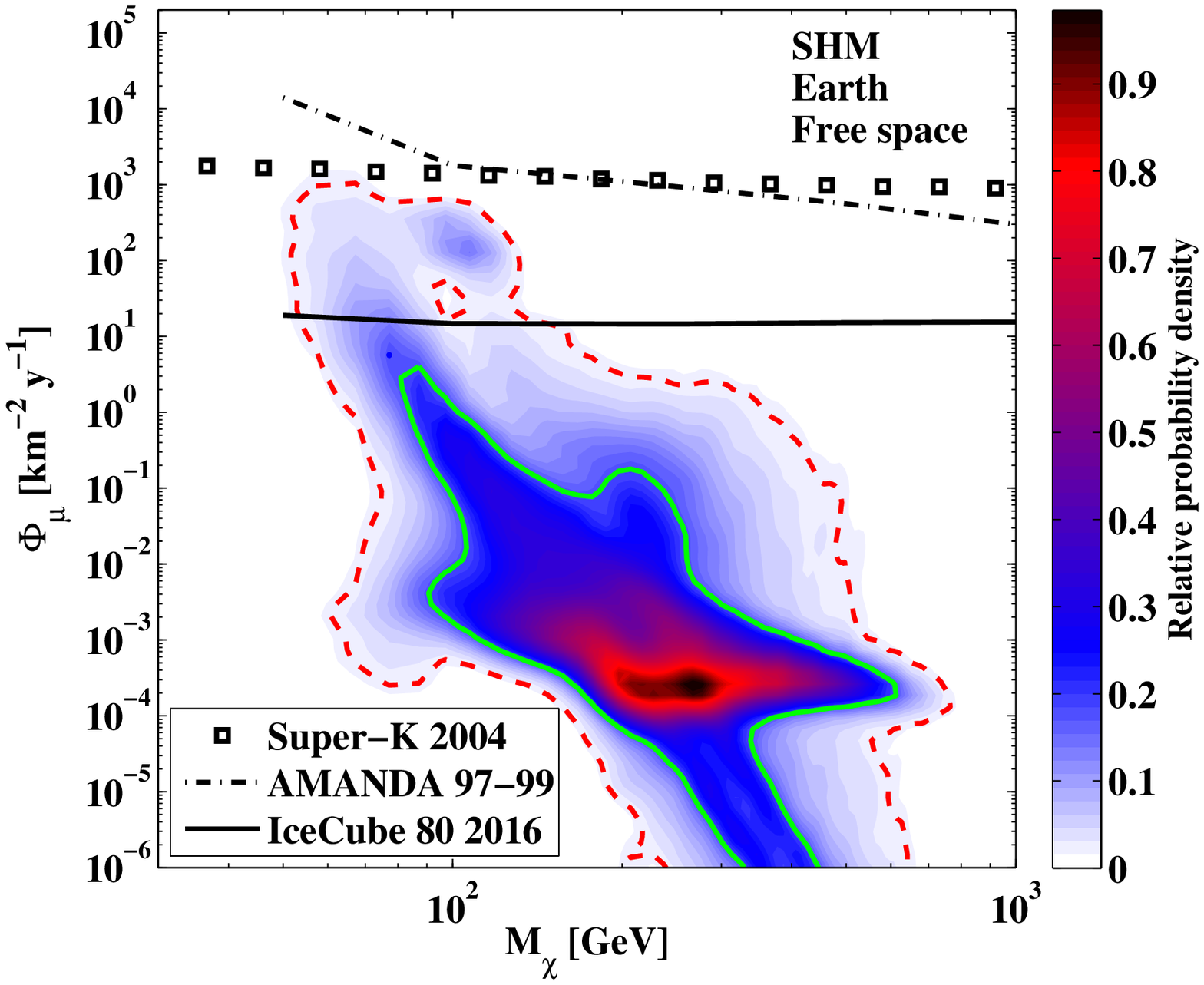} 
 \includegraphics*[width=0.45\textwidth]{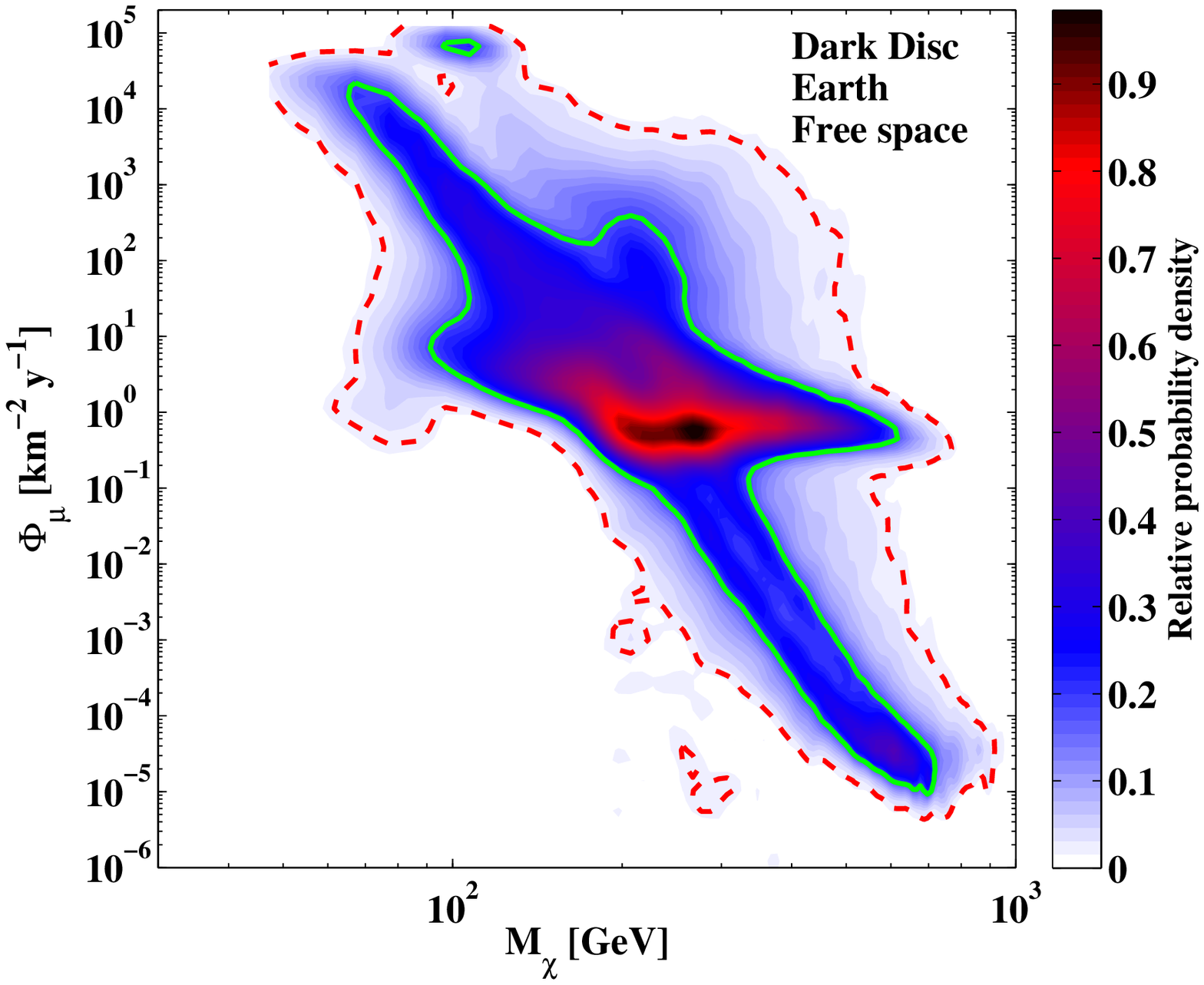}
 \includegraphics*[width=0.45\textwidth]{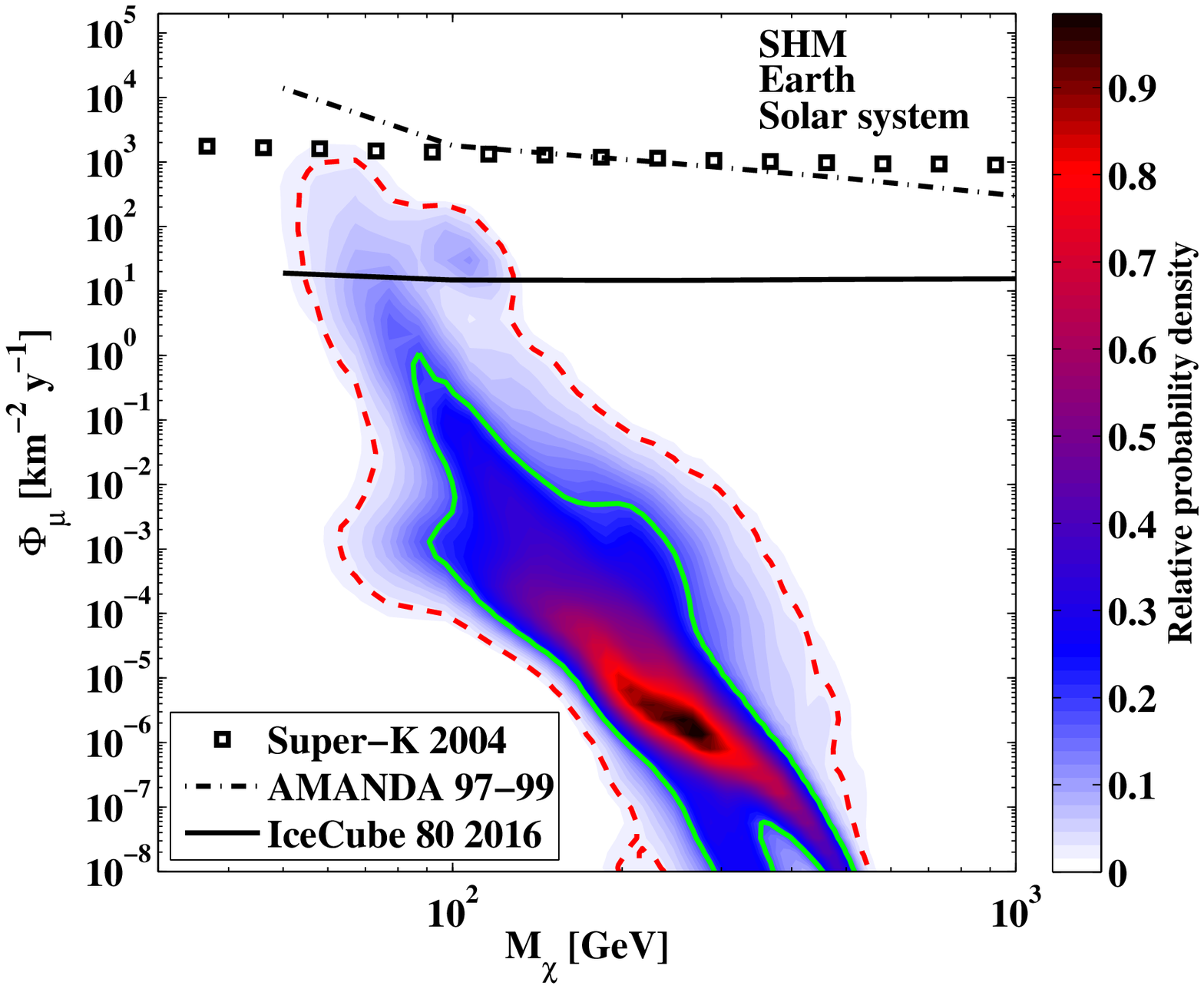} 
 \includegraphics*[width=0.45\textwidth]{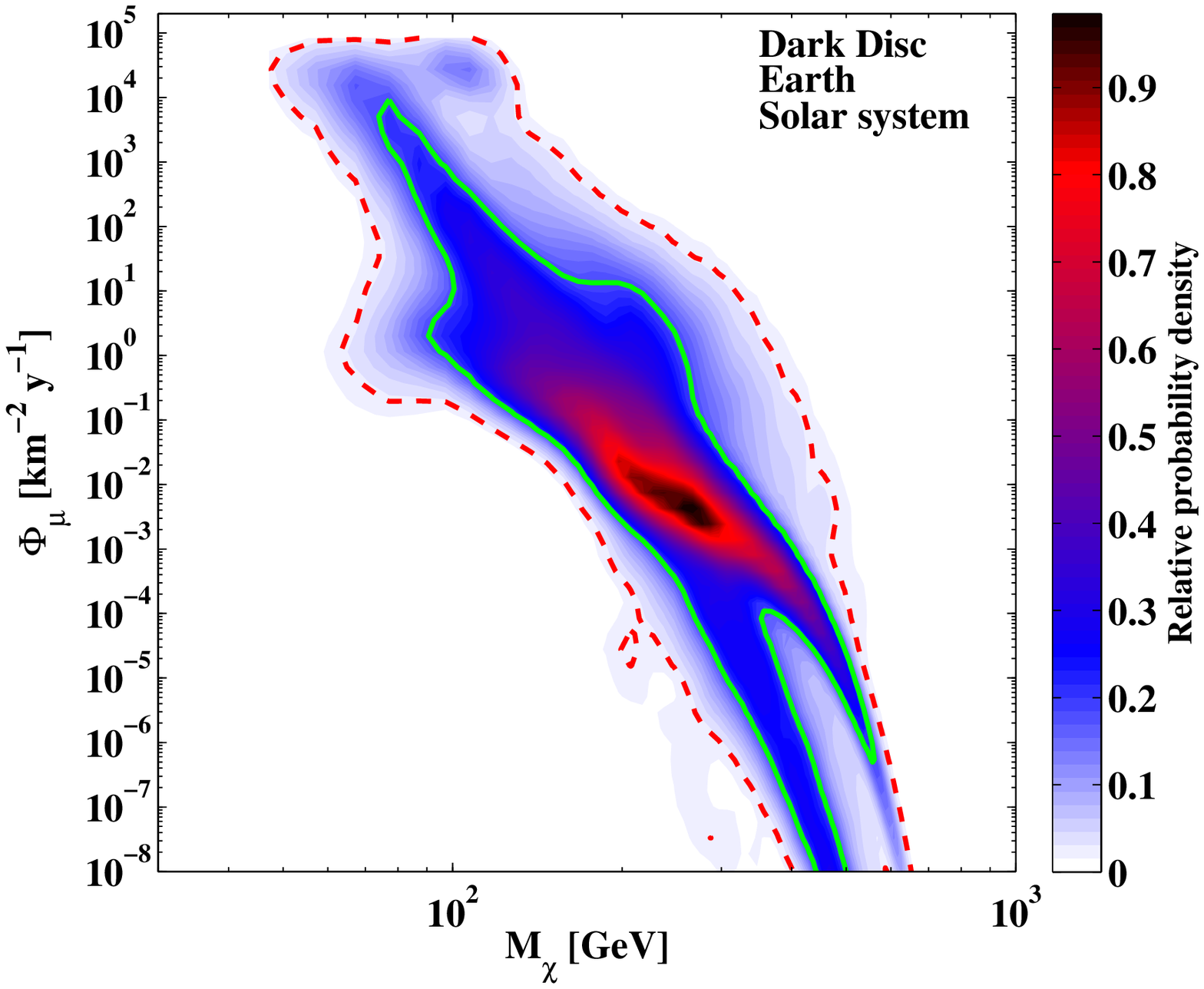}
\label{fig:earthflux}
\end{center}
\caption{Muon flux $\Phi_{\mu}$ for $E_{\mu} > 1$\,GeV at the Earth's surface as a function of $\mwimp$ from neutrinos originating in the Earth. The top row is calculated with the free space Gaussian approximation of the velocity distribution, while for the bottom row the velocity distribution from the Solar system simulations is used. The sharp decrease in the muon flux at high WIMP masses in the bottom row is caused by the kinematic cut-off of the capture rate (see text). Compared to flux from the SHM (left panels) the flux from the dark disc (right panels) is boosted by two to three orders of magnitude for $\rhodrat=1$, depending on the specific model. Current experimental constraints on the muon flux from the Earth from Super-Kamiokande \cite{SuperK2004} and AMANDA-II \cite{amanda7yr, amandab10} along with the expected sensitivity of IceCube80 are compared to the flux expected from the SHM. The closed contours show -- 95\% (red/dashed) and 68\% (green/solid) -- of the probability density of CMSSM models consistent with both astrophysical and collider constraints, and assuming flat priors. The colour-bar gives the relative probability density (see \S\ref{sec:dm} for details). \textit{Note the vertical scales of the two bottom plots differ by two orders of magnitude as compared with the top plots.}}
\end{figure*}

\subsection{The Earth}
Finding the muon flux from WIMP annihilation in the Earth is somewhat more complicated.  Because the escape velocity of the Earth is small ($v \approx 15$\,km/s at the centre), capture is only possible for low speed WIMPs unless the WIMP mass is nearly identical to that of one of the nuclear species in the Earth \cite[Eq. \ref{eq:umax}; see also ][]{Gould1987,Gould1988}.  Moreover, the capture rate is disproportionately sensitive to the lowest speed WIMPs since those WIMPs may be captured anywhere in the body, whereas higher speed WIMPs may only be captured at the centre where the escape velocity is largest.  However, the low speed tail of the WIMP speed distribution is not precisely known; for speeds relative to the Earth of $u < 72 $\,km/s (the speed of a WIMP at the escape velocity from the Solar system, moving in the direction opposite to the Earth), the phase space may be occupied by WIMPs bound to the Solar system as well as Galactic WIMPs streaming through the Solar system on unbound orbits.  Thus, the annihilation rate of WIMPs in the Earth depends on the density of WIMPs bound to the Solar system, which has not yet been definitively determined.

\begin{figure*}[!thb]
\begin{center}
 \includegraphics[width=0.45\textwidth]{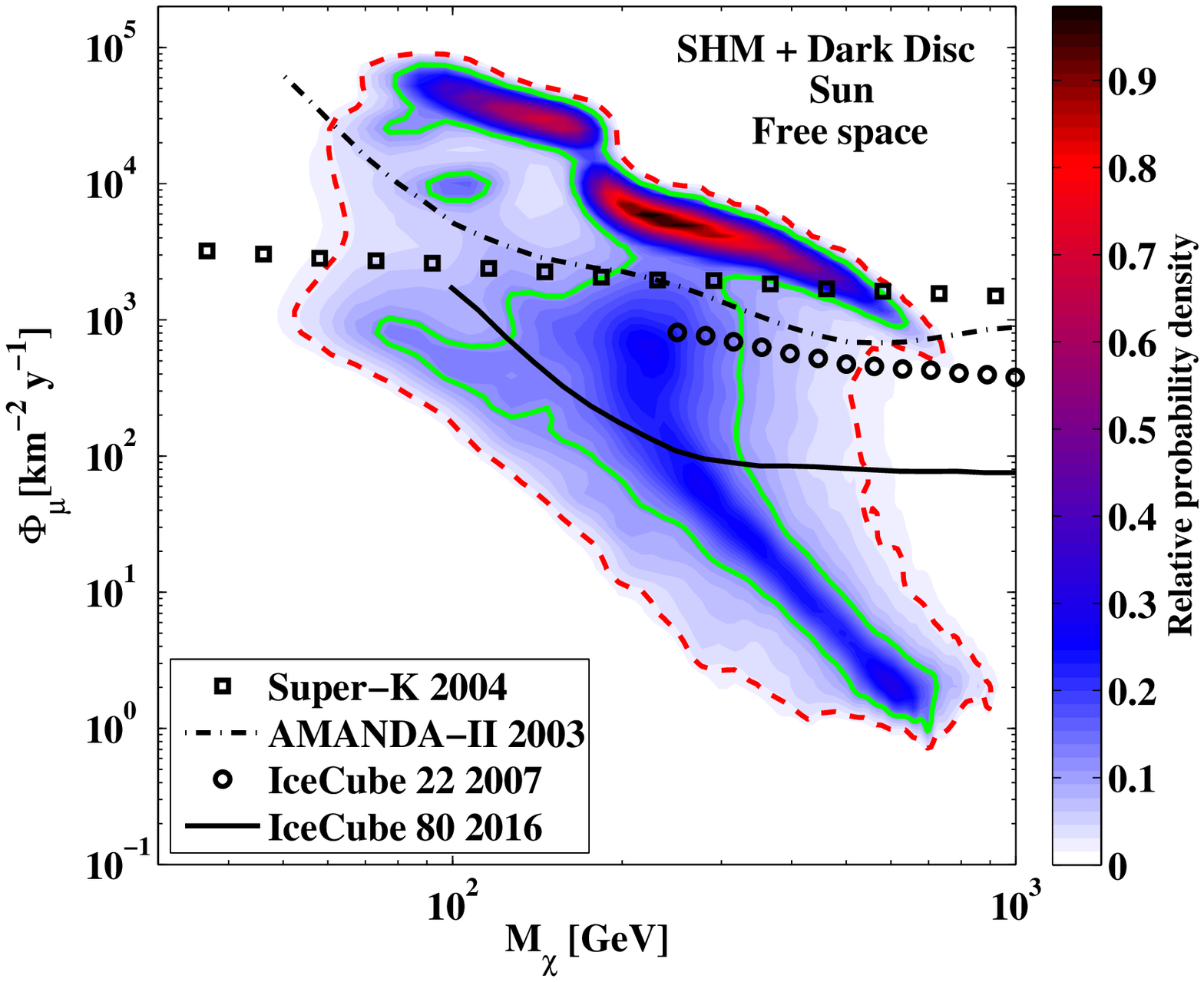} 
 \includegraphics[width=0.45\textwidth]{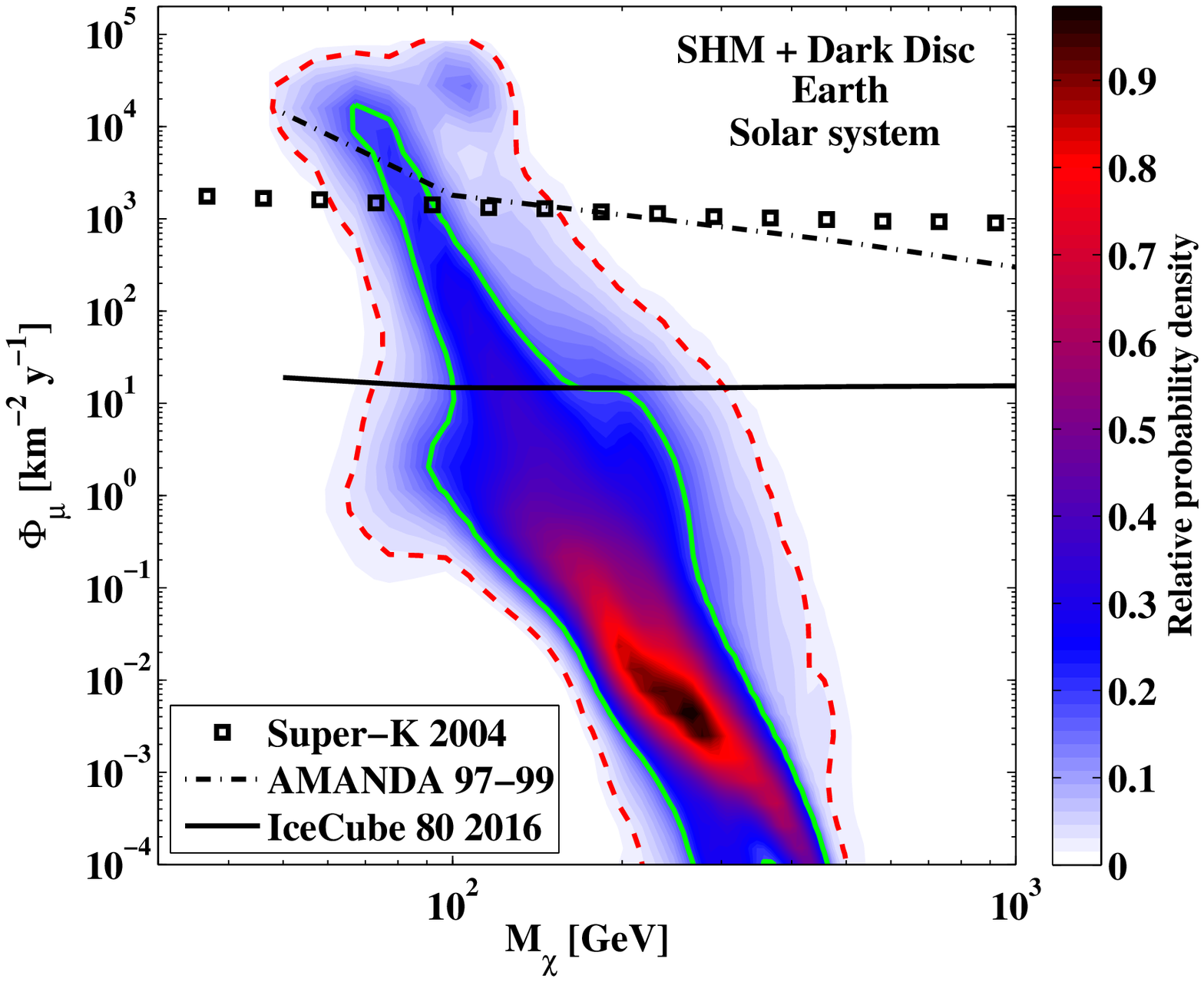} 
\label{fig:totflux}
\end{center}
\caption{Total muon flux $\Phi_{\mu}$ for $E_{\mu} > 1$\,GeV at the Earth's surface as a function of $\mwimp$ from neutrinos originating in the Sun (left panel) and Earth (right panel). In both cases the muon flux is dominated by the dark disc component. Current experimental constraints on the muon flux from the Earth and Sun from Super-Kamiokande \cite{SuperK2004}, AMANDA-II \cite{amanda7yr, amandab10} and IceCube22 \cite{icecube22} are shown. The enhanced flux allows these experiments to constrain a much larger portion of the CMSSM parameter space. The projected sensitivity of the IceCube80 experiment will probe a significant fraction of the allowed parameter space. The closed contours show -- 95\% (red/dashed) and 68\% (green/solid) -- of the probability density of CMSSM models consistent with both astrophysical and collider constraints, assuming flat priors. The colour-bar gives the relative probability density (see \S\ref{sec:dm} for details).}
\end{figure*}

\subsection{Sensitivity to the population bound to the Solar system}
There is a spread in predicted bound WIMP distributions. Although the following studies predicted the bound WIMP distribution for the SHM, the results generalise to arbitrary dark matter distributions. On the high end, \citet{Gould1991} argued that the low speed WIMP distribution resulting from gravitational capture of WIMPs by the planets should be approximately the free space Gaussian distribution function of Eq. (\ref{eq:gauss}).  This argument was based on treating WIMP-planet encounters as local, with the cumulative changes to WIMP speed treated in the random walk approximation.  Also using the local approximation, \citet{Lundberg2004} found a smaller low speed WIMP distribution if they treated the Sun as being infinitely optically thick to WIMPs.  \citet{Damour1999} considered a population of long-lived WIMPs captured in the Solar system by elastic scattering in the Sun, but neglected subsequent scatters of those WIMPs with solar nuclei.  \citet{Bergstrom1999} found that this population could boost the annihilation rate of WIMPs in the Earth by a factor of $\sim 100$ for $60 \hbox{ GeV} < \mwimp < 130 \hbox{ GeV}$.

More recently, Peter \cite{Peter2009a, Peter2009b} has simulated $\sim 10^{10}$ WIMPs bound to the Solar system by either gravitational capture or elastic scattering in the Sun.  Orbits were integrated in a toy Solar system consisting of the Sun and Jupiter.  WIMP trajectories were followed using a modified symplectic integration algorithm, allowing for the possibility of further elastic scattering in the Sun. The orbits were integrated until the WIMPs were ejected or scattered onto orbits that no longer intersected the Earth. The phase space density distribution of bound WIMPs as a function of speed relative to the Earth is shown in Fig. 2 for both the SHM and the dark disc.  Also shown in Fig. 2 are the phase space density distributions of only the Galactic WIMPs (unbound to the Solar system) and the free space Gaussian approximation (denoted as ``Free space'' on relevant figures).

The bound velocity distribution is significantly smaller than predicted by \citet{Gould1991} and \citet{Damour1999}, and similar to that found by \citet{Lundberg2004}.  While part of the difference is due to elastic scattering in the Sun (especially for the Damour and Krauss population), part of the difference is due to simulating orbits in a toy Solar system. The cut-off in the velocity distribution at $u \sim 9$\,km/s owes to the phase space below being inaccessible to WIMPs in the toy Solar system due to the conservation of the Jacobi integral of motion; interaction with the inner planets is required to populate lower speeds.  This cut-off in speed translates to a cut-off in muon flux above a particular WIMP mass.  Solving Eq. (\ref{eq:umax}) for $M_\chi$, and setting $m_i$ to the mass of $^{56}$Fe (the dominant atomic species in the core of the Earth), we find that the muon flux is exactly zero for $M_\chi > 700$ GeV.  The impact of the WIMPs bound to the Solar system on the WIMP annihilation rate in the Earth depends crucially on how effective the inner planets are at populating the phase space below $u = 9$\,km/s.

Given the uncertainty in the low speed WIMP distribution, we calculate annihilation rates for both the Peter \cite{Peter2009a, Peter2009b} distribution functions and the free space Gaussian distribution function.  These span a likely range of the true distribution function of low speed WIMPs.  The \citet{Damour1999} solar-captured distribution function is larger than the Gaussian distribution function for $30\hbox{ km/s} < u < 50 \hbox{ km/s}$. However Peter \cite{Peter2009a}  finds that subsequent scattering in the Sun reduces the lifetime and phase space density of these WIMPs below the Gaussian phase space density.

In Fig. 3, we show the muon fluxes from WIMP annihilation in the Earth for both the SHM and the dark disc assuming a muon energy threshold of 1\,GeV.  For both distribution functions, the flux from the dark disc is two to three orders of magnitude above the SHM if $\rhodrat=1$.  This large increase is due to the fact that $t_\odot/\tau < 1$, such that $\Gamma \propto C^2$.  Thus, an increase in the capture rate of WIMPs in the Earth has a more dramatic effect on the muon flux than a similar enhancement in the capture rate of WIMPs in the Sun.  

The predicted flux from WIMPs with $\mwimp\gtrsim100$\,GeV is quite sensitive to the low speed phase space density distribution.  For the distribution function from the Solar system simulations, we find the steep drop in flux due to the kinematic cut-off in the capture rate for $\mwimp\gtrsim500$\,GeV.  As a consequence, while the enhancement of the muon flux from the dark disc puts the search for WIMP annihilation in the Earth on the same level as the Sun for $\mwimp \lesssim 100\hbox{ GeV}$, the prospects for detecting WIMPs of higher masses is unclear. Precision estimates of the low speed tail of the WIMP velocity distribution are necessary to determine the prospects for high mass WIMPs.

\section{Discussion}\label{sec:discussion}

In Fig. 4, we show the total flux from the Sun and the Earth (including capture from both SHM and dark disc components assuming $\rhod/\rhoh = 1$) along with current experimental constraints. The flux in both cases is dominated by the dark disc component. To be conservative, we show the lower bound of the expected muon flux from the Earth obtained using the phase space density distribution from the Solar system simulations. The inclusion of the dark disc component significantly improves the constraints on the allowed parameter space from current experiments.  Large area neutrino telescopes such as IceCube will be sensitive to a large fraction of the allowed parameter space, providing a complementary search for dark matter to direct detection experiments.

Systematic uncertainties owing to the unknown density and velocity distribution of the dark disc are especially large for the Earth owing to high powers of these parameters in the calculation
of the annihilation flux. For the results presented in this Letter, we used $\rhodrat=1$ and $\sigma_d = 50$\,km/s with the mean lag $|\mathbf{v}_\odot| = \sigma_d$.
For the Earth the dependency is given by $\exp(-|\mathbf{v}_\odot|^2/\sigma_d^2)(\rhodrat)^2/\sigma_d^{6}$ for masses $\mwimp>100$\,GeV, since the part of the WIMP phase space relevant for capture scales as $\exp(-|\mathbf{v}_\odot|^2/(2\sigma_d^2))(\rhodrat)/\sigma_d^{3}$
and the flux depends on the capture rate squared.  For the Sun the actual dependency is more complex \footnote{The  capture rate scales as \begin{equation}\frac{(\rhodrat)}{|\mathbf{v}_\odot|} \left(2\cdot \mathrm{erf} \left(\frac{|\mathbf{v}_\odot|}{\sqrt{2} \sigma_d}\right)-\mathrm{erf}\left(\frac{|\mathbf{v}_\odot|- v_{cut}}{\sqrt{2} \sigma_d}\right)- \mathrm{erf}\left(\frac{|\mathbf{v}_\odot|+ v_{cut}}{\sqrt{2} \sigma_d}\right)\right) \end{equation}\noindent where $v_{cut} \sim 2000 \sqrt{GeV/\mwimp}$\,km/s is approximately the maximum speed of WIMPs which can be captured in the Sun.}. An empirical estimate for $|\mathbf{v}_\odot| = \sigma_d$ of the dependency gives a scaling of $(\rhodrat)/\sigma_d^a$ with $a \in [1~2]$ depending on the particle's mass. This scaling differs from the scaling in the Earth owing to the flux being proportional to one power of the capture rate. Simulations have shown that all disc galaxies will have a dark disc, but the cosmic variance in its
properties will be large.  At the minimum extreme is a dark disc with $\rhodrat = 0.25$ and $\sigma_d \simeq 100$\,km/s (corresponding to the lowest/highest value respectively, found in the simulations \cite{Read09}). Even in this case, the annihilation signal from the Earth and Sun are both dominated by the dark disc rather than the dark halo.  However, with such a large velocity dispersion, the scaling just described means that the dark disc does not lead to the large boosts that come from our median dark disc properties.

The median values for the dark disc properties exclude the most probable regions of the CMSSM parameter space. However, given the uncertainties in the dark disc properties we cannot yet convincingly exclude relevant CMSSM parameter space shown in Fig. 4.

Future surveys of our Galaxy like RAVE \cite{Rave06} and GAIA \cite{GAIA} will detect the local density of dark matter and may disentangle accreted stars (which will have nearly the same velocity dispersion as the dark disc) from those formed in-situ. In this case, it will be possible to infer the actual properties of the dark disc from these stars, and hence, make more robust predictions for the event rate in neutrino telescopes.

\section{Conclusions}\label{sec:conclusions}

In \lcdm, a dark matter disc forms from the accretion of satellites in disc galaxies.  We have shown how its increased phase space density at low velocities  enhances the capture rate of dark matter particles in the Earth and Sun, resulting in an increased neutrino-induced muon flux at the Earth from WIMP annihilation. Our main findings are: 

\begin{enumerate}

\item The dark disc significantly boosts the capture rate of dark matter particles in the Sun and Earth as compared to the SHM. This increase owes to the higher phase space density at low velocities in the dark disc. For the Sun, the expected muon flux from the dark disc with $\rhodrat=1$ is increased by one order of magnitude relative to a pure SHM-generated flux.  If the WIMP is the neutralino in the CMSSM, neutrino telescopes will explore a large fraction of the CMSSM parameter space.  

\item For the Earth --- where WIMP capture and annihilation are not in equilibrium --- the increase in the muon flux is two to three orders of magnitude, although this depends sensitively on the distribution function of the dark disc.  
For the SHM alone, the flux from the Sun is far greater than that from the Earth.
The enhancement from the dark disc puts the search for WIMP annihilation in the Earth on the same level as the Sun if $\mwimp \lesssim 100\hbox{ GeV}$\footnote{Our particular WIMP model already has little parameter space below 100 GeV, but this extra sensitivity of detection below 100 GeV owes to kinematics of Solar system transport and capture, not the WIMP model.}. For larger WIMP masses, the prospects for detecting muons from annihilation in the Earth requires better models of the density of WIMPs bound to the Solar system.  

\end{enumerate}

We thank the IceCube Collaboration for providing their expected sensitivities. We would like to thank Ivone F.\,I.~Albuquerque for fruitful initial discussions on this study.
We acknowledge support from the Swiss NSF and the wonderful working environment and support
of UZH. Annika H.\,G.~Peter acknowledges support from the Gordon and Betty Moore Foundation.

\end{document}